\documentclass[aps,prb,reprint,superscriptaddress,showpacs,floatfix]{revtex4-1}

\usepackage{graphicx}
\usepackage{amssymb,amsmath}
\usepackage{dcolumn}
\usepackage{bm}
\usepackage[mathlines]{lineno}

\begin{document}

\title{\hfill {\small Phys. Rev. B (2014)}\\
       Local curvature and stability of two-dimensional systems
       }

\author{Jie Guan}
\affiliation{Physics and Astronomy Department,
             Michigan State University,
             East Lansing, Michigan 48824, USA}

\author{Zhongqi Jin}
\affiliation{Chemistry Department,
             Michigan State University,
             East Lansing, Michigan 48824, USA}

\author{Zhen Zhu}
\affiliation{Physics and Astronomy Department,
             Michigan State University,
             East Lansing, Michigan 48824, USA}

\author{Chern Chuang}
\affiliation{Department of Chemistry,
             Massachusetts Institute of Technology,
             Cambridge, MA 02139, USA}

\author{Bih-Yaw Jin}
\affiliation{Department of Chemistry and
             Center for Emerging Material and Advanced Devices,
             National Taiwan University,
             Taipei 10617, Taiwan}

\author{David Tom\'{a}nek}
\email
            {tomanek@pa.msu.edu}%
\affiliation{Physics and Astronomy Department,
             Michigan State University,
             East Lansing, Michigan 48824, USA}

\date{\today} 

\begin{abstract}
We propose a fast method to determine the local curvature in
two-dimensional (2D) systems with arbitrary shape. The curvature
information, combined with elastic constants obtained for a planar
system, provides an accurate estimate of the local stability in
the framework of continuum elasticity theory. Relative stabilities
of graphitic structures including fullerenes, nanotubes and
schwarzites, as well as phosphorene nanotubes, calculated using
this approach, agree closely with {\em ab initio} density
functional calculations. The continuum elasticity approach can be
applied to all 2D structures and is particularly attractive in
complex systems with known structure, where the quality of
parameterized force fields has not been established.
\end{abstract}



\maketitle

\section{Introduction}

Layered structures including graphite, hexagonal boron nitride,
black phosphorus, transition metal dichalcogenides such as
MoS$_2$, and oxides including V$_2$O$_5$ are very common in
Nature. The possibility to form stable two-dimensional (2D)
structures by mechanical exfoliation of these structures appears
very attractive for a variety of
applications.\cite{{Novoselov2005pnas},{Feldman1995science}} The
most prominent example of such 2D systems, graphitic carbon, is
the structural basis not only of graphene,\cite{Novoselov2005pnas}
but also
fullerenes, nanotubes, tori and schwarzites.%
\cite{{Kroto85},{Iijima91},{Chuang09},{schwarzite91},{DT227}} Even
though the structural motif in all of these systems may be the
same, their mechanical and electronic properties depend
sensitively on the local
morphology.\cite{{Lu2005cr},{Pereira09},{cnt-deformation}}
Not only does the natural abundance of structural allotropes and
isomers reflect their net energetic stability, but also the
relative chemical reactivity of specific sites in a given
structure correlates well with the local curvature and local
stability.\cite{{Lu2005cr},{Pereira09},{cnt-deformation}} This
relationship has been well established for the reactive sites in
the C$_{50}$ fullerene,\cite{Lu2005cr} used to induce structural
collapse leading to chemical unzipping of carbon
nanotubes,\cite{{DT200},{Liying2009nature},{Dmitry2009nature}} and
to destroy collapsed carbon nanotubes.\cite{cnt-deformation}

\begin{figure}[!b]
\includegraphics[width=0.9\columnwidth]{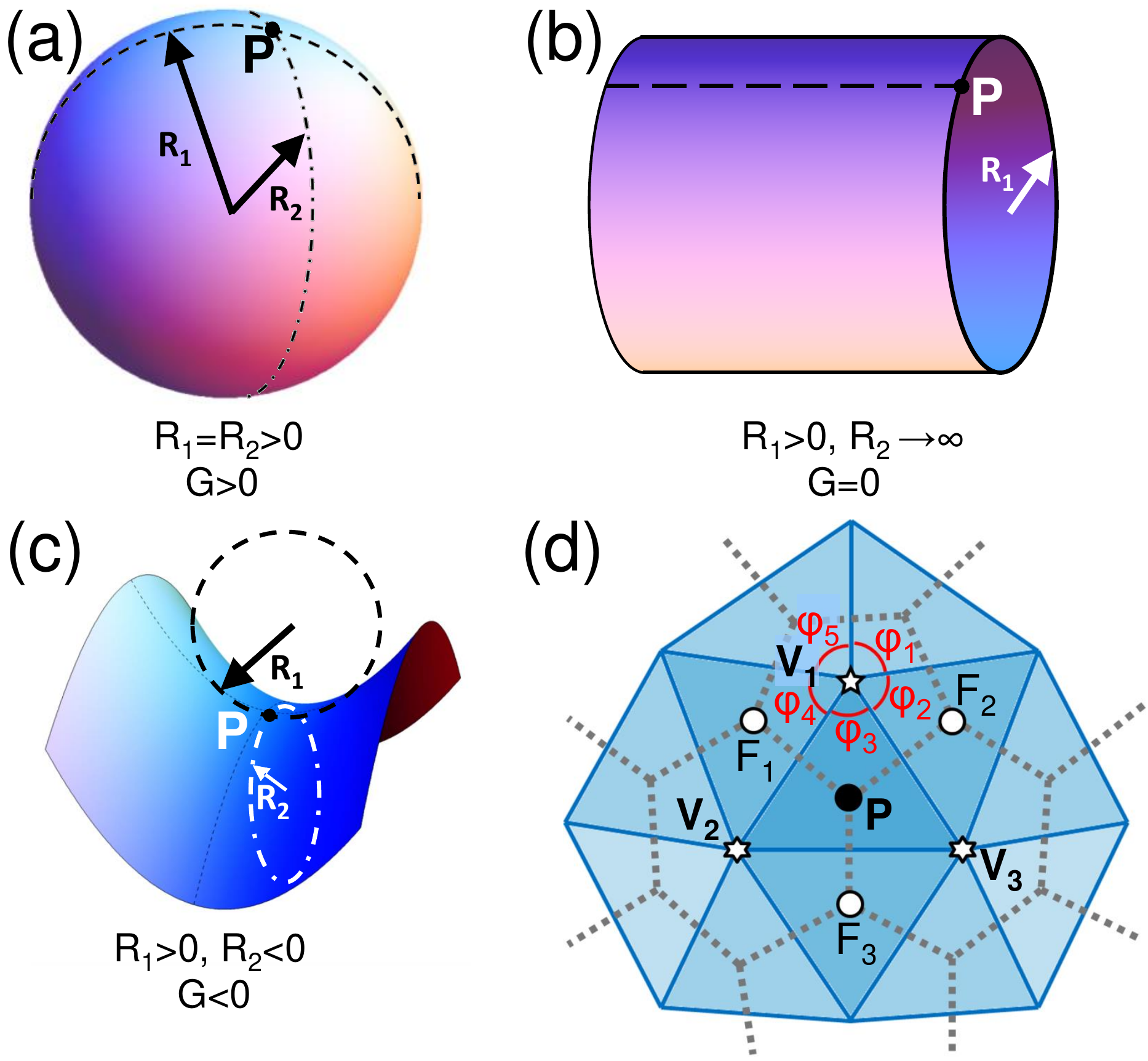}
\caption{%
(Color online) Principal radii of curvature $R_1, R_2$ and the
Gaussian curvature $G$ on the surface of (a) a sphere, (b) a
cylinder and (c) in a saddle point. (d) Determination of the local
curvature at point $P$ using the atomic lattice and the dual
lattice. \label{fig1} }
\end{figure}

For very large structures, estimating the global or local
stability using {\em ab initio} calculations has proven
impracticable. There, the stability has often been estimated using
empirical rules or parameterized force fields
including the Tersoff potential and molecular mechanics,%
\cite{{Tersoff88},{Albertazzi1999pccp},{Schmalz1988jacs},{Kroto1987nature}}
with sometimes unsatisfactory results. Application of continuum
elasticity theory, which can describe stability changes due to
deviation from planarity, has been successful, but limited to
systems with a well-defined, constant
curvature.\cite{{DT071},{Enyashin2007epj}} Since strain energy is
local and independent of the global morphology, it is intriguing
to explore,
whether the local deformation energy may be accurately determined
from local morphology estimates using the atomic geometry. If so,
then the local stability in even arbitrarily shaped structures
could be estimated accurately.

Here we propose a fast method to determine the local curvature in
2D systems with a complex morphology using the local atomic
geometry. Curvature information alone, combined with elastic
constants obtained for a planar system, provides accurate
stability estimates in the framework of continuum elasticity
theory. We find that relative stabilities of graphitic structures
including fullerenes, nanotubes and schwarzites, as well as
phosphorene nanotubes, calculated using this approach, agree
closely with {\em ab initio} density functional calculations. The
continuum elasticity approach can be applied to all 2D structures
and is particularly attractive in complex systems with known
structure, where the quality of parameterized force fields has not
been established.

\section{Local curvature and curvature energy}\label{Sec2}

The local curvature at a particular location on a surface is given
by the two principal radii of curvature $R_1$ and $R_2$, as shown
in Fig.~\ref{fig1}. On a spherical surface, $R_1=R_2$. On a
cylindrical surface, $R_1$ is the cylinder radius and
$R_2{\rightarrow}\infty$. Finally, a saddle point on a surface is
characterized by opposite signs of $R_1$ and $R_2$. Knowing the
principal radii of curvature everywhere, we may use continuum
elasticity theory to determine the curvature energy
${\Delta}E_{C}$ with respect to a planar layer
using\cite{Maceri2010}
\begin{equation}
{\Delta}E_{C} = \frac{1}{2} D \int_{surface} dA
        \left(
        \frac{1}{R_1^2} + \frac{1}{R_2^2} + \frac{2\alpha}{R_1R_2}
        \right) \;.
\label{eq1}
\end{equation}
Here, the integral extends across the entire closed surface, $D$
is the flexural rigidity and $\alpha$ is the Poisson ratio. Simple
expressions for ${\Delta}E_{C}$ can be obtained for
simple morphologies such as a sphere or a cylinder, where $R_1$
and $R_2$ are constant everywhere.\cite{DT071} This is, however,
not the case in general.

We find it convenient to introduce the local mean curvature
\begin{equation}
k = \frac{1}{2} \left( \frac{1}{R_1} + \frac{1}{R_2} \right)
\label{eq2}
\end{equation}
and the local Gaussian curvature
\begin{equation}
G = \frac{1}{R_1R_2} \;. %
\label{eq3}
\end{equation}
Using these quantities, we can rewrite Eq.~(\ref{eq1}) as
\begin{equation}
{\Delta}E_{C} = D \int_{surface} dA
        \left[ 2k^2 - (1-{\alpha}) G \right] \;.
\label{eq4}
\end{equation}
%

In the following, we will consider the equilibrium arrangement of
atoms in a planar 2D structure as the reference structure and will
determine the local curvature from changes in the local
morphology. The discrete counterpart of Eq.~(\ref{eq4}) for the
curvature energy ${\Delta}E_{C}$ is a sum over atomic sites $i$,
\begin{equation}%
{\Delta}E_{C} {\approx} D A \sum_{i}
    \left[ 2k_i^2 - (1-{\alpha}) G_i \right] \;, %
\label{eq5}
\end{equation}%
where $A$ is the area per atom.

To use Eq.~(\ref{eq5}) for curvature energy estimates, we need to
know the local curvatures $k$ and $G$ at all atomic sites. Our
approach to estimate these values at a given site $P$ is
illustrated in Fig.~\ref{fig1}(d). According to Eq.~(\ref{eq2}),
the local mean curvature $k$ should be close to the average
inverse radius of curvature at that point,
\begin{equation}%
k {\approx} \left< \frac{1}{R} \right> \;. %
\label{eq8}
\end{equation}%
Since the atomic site $P$ and its nearest three neighbors $F_1$,
$F_2$ and $F_3$ define the surface of a sphere of radius $R$, we
take $k=1/R$.

The positions of four atoms do not allow to distinguish, whether
$P$ is on a plane, a sphere, a cylinder, or in a saddle point. We
may obtain this additional information using the concept of
angular defect. On any surface, which can be triangulated as shown
in Fig.~\ref{fig1}(d), the angular defect at a representative
vertex $V_1$ is defined by
$\Delta(V_1)=2\pi-\sum_i\varphi_i$
in radian units. The local Gaussian curvature at $V_1$ is then
given by\cite{Meek2000}
\begin{equation}
G(V_1)=\Delta(V_1)/A_t = \left(2\pi-\sum_i\varphi_i\right)/A_t \;,%
\end{equation}
where $A_t$ is the total area of the triangulated surface divided
by the number of vertices.

For trivalent molecular graphs containing 5-, 6- and 7-membered
rings found in fullerenes, carbon nanotubes and schwarzites, a
unique triangulation may be obtained by connecting the centers of
adjacent polygons. This method is referred to as the dual graph in
graph theory\cite{ChuangDual09} and its use is illustrated in
Fig.~\ref{fig1}(d). Since $P$ is not a vertex in the dual graph,
but rather the center of the triangle $\bigtriangleup V_1V_2V_3$,
we must infer the local Gaussian curvature at $P$ from the angular
defects at $V_1$, $V_2$ and $V_3$. If vertex $V_j$ is surrounded
by $n_j$ triangles, we may assign to point $P$ the angular defect
$\Delta(P)= \Delta(V_1)/n_1+\Delta(V_2)/n_2+\Delta(V_3)/n_3$. %
Then, we can estimate the local Gaussian curvature at $P$ as
\begin{equation}
G(P)=\Delta(P)/A \;,
\end{equation}
where $A$ is the average area per atom. We use $A=2.62$~{\AA}$^2$,
the value found in the honeycomb lattice of graphene, for all
graphitic structures.

The above definition of the local Gaussian curvature satisfies
exactly the equality
\begin{equation}
A\sum_{atoms} G(P_j) = A_t \sum_{vertices} G(V_j) = 2\pi\chi \;. %
\label{eq11}
\end{equation}
Here, $\chi$ is the Euler characteristic of the surface, given by
%
$\chi = 2 - 2g$,
%
where $g$ is the genus, meaning `number of holes'. Of interest
here is the fact that $\chi=2$ for spherical objects like
fullerenes and $\chi=0$ for cylindrical objects such as nanotubes.
Equation (\ref{eq11}) is the discretized version of the
Gauss-Bonnet theorem\cite{Encyclopedia-of-math} regarding the
integral of the Gaussian curvature over an entire closed surface,
called the sum of the defect, which is usually formulated as
$\int_{surface} G dA = 2{\pi}{\chi}$.

\begin{figure}[t]
\includegraphics[width=0.85\columnwidth]{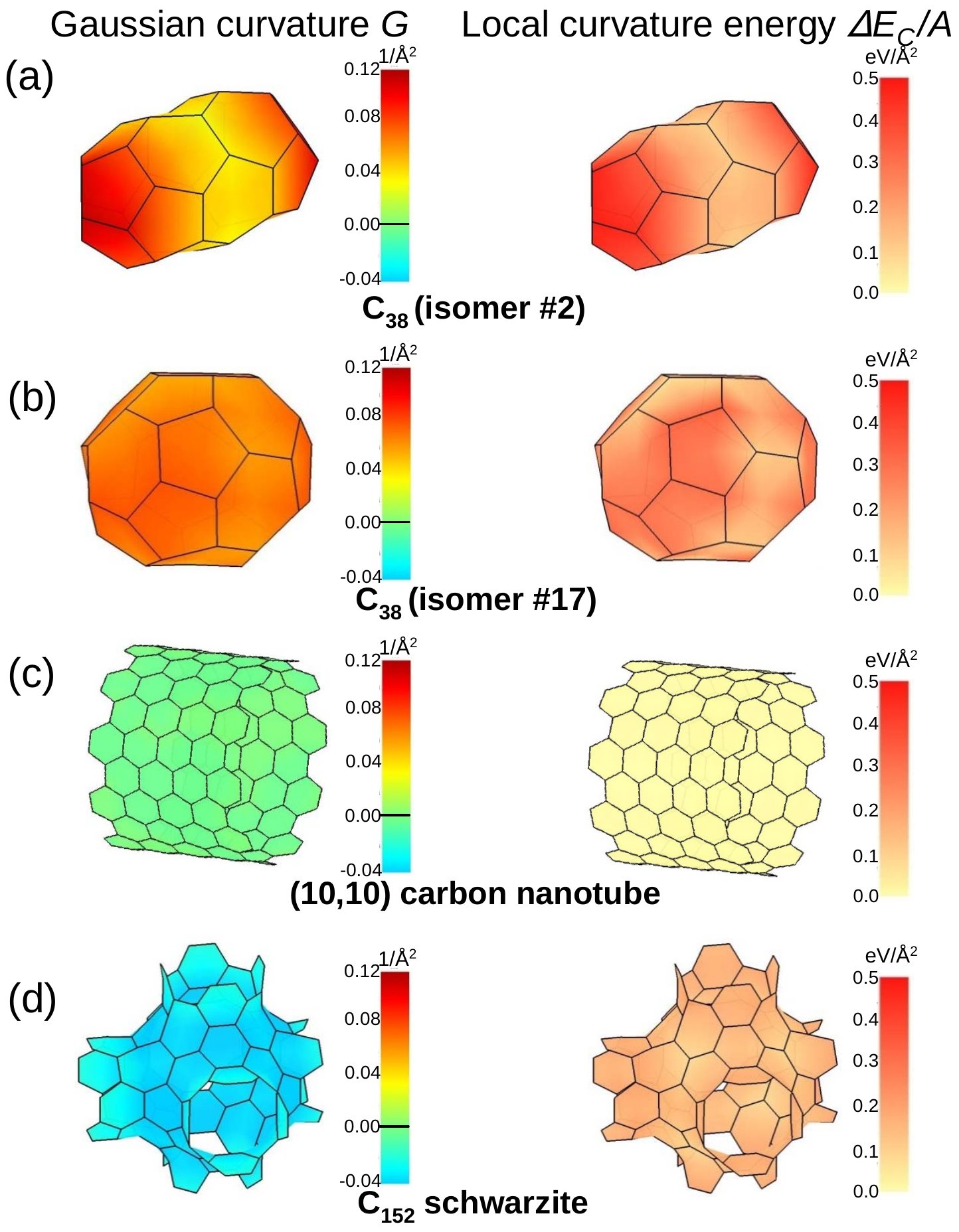}
\caption{%
(Color online) Local Gaussian curvature $G$ (left panels) and
local curvature energy ${\Delta}E_{C}/A$ across the surface of (a)
the least stable C$_{38}$ isomer, (b) the most stable C$_{38}$
isomer, (c) a (10,10) carbon nanotube, and (d) a schwarzite
structure with 152 atoms per unit cell. The values of $G$ and
${\Delta}E_{C}/A$ have been interpolated across the surface.
\label{fig2} }
\end{figure}

The variation of the local Gaussian curvature $G$ and the local
curvature energy ${\Delta}E_C/A$ across the surface of carbon
polymorphs, including two fullerene isomers discussed in
Section~\ref{Sec3}.B, a nanotube and a schwarzite structure, is
displayed in Fig.~\ref{fig2}. The local curvature energy in these
$sp^2$-bonded structures has been evaluated using the elastic
constants of graphene\cite{DT071} $D=1.41$~eV and $\alpha=0.165$.
The higher stability of the C$_{38}$(17) isomer in
Fig.~\ref{fig2}(b) is reflected in a rather uniform curvature
energy and Gaussian curvature distribution. The low stability of
the C$_{38}$(2) isomer in Fig.~\ref{fig2}(a) is reflected in a
large variation of curvature energy and Gaussian curvature,
clearly indicating the most reactive sites. Cylindrical carbon
nanotubes, such as the (10,10) nanotube displayed in
Fig.~\ref{fig2}(c), have zero Gaussian curvature and a constant
curvature energy caused by the mean curvature. Schwarzites such as
the C$_{152}$ structure, displayed in Fig.~\ref{fig2}(d), have
only negative Gaussian curvature that may vary across the surface,
causing variations in the local curvature energy.

\begin{figure*}[!t]
\includegraphics[width=1.8\columnwidth]{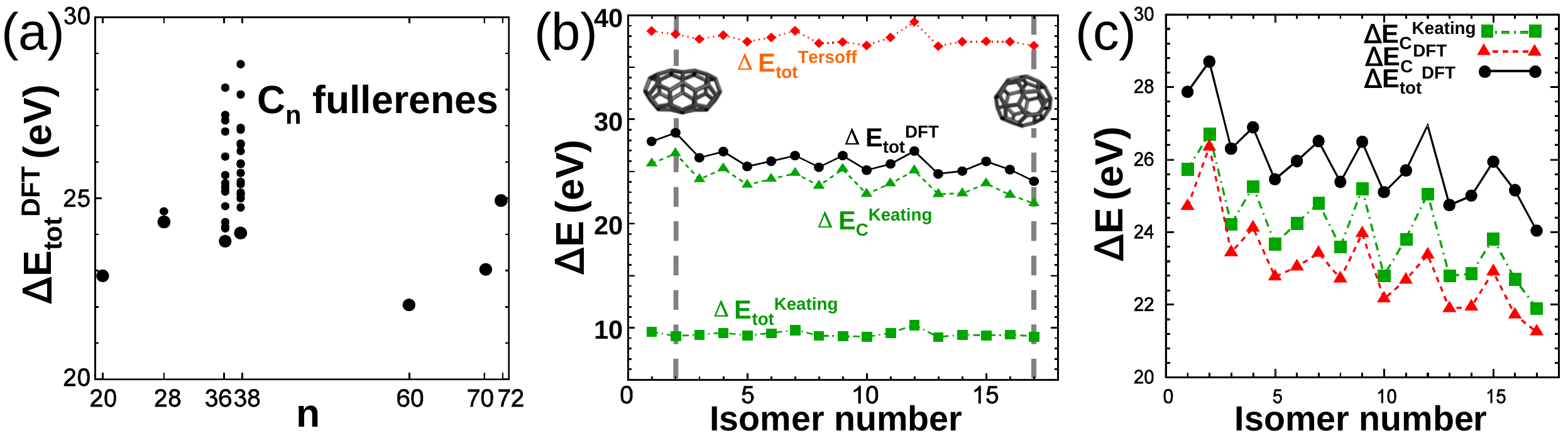}
\caption{%
(Color online) Strain energy ${\Delta}E$ in carbon nanostructures
with respect to the graphene reference system. (a) DFT-based total
strain energy ${\Delta}E_{tot}^{DFT}$ for selected fullerenes,
with the most stable isomers indicated by the larger symbols. (b)
Strain energy ${\Delta}E$ in different C$_{38}$ isomers. Total
energy differences ${\Delta}E_{tot}^{DFT}$ based on DFT,
${\Delta}E_{tot}^{Tersoff}$ based on the Tersoff potential and
${\Delta}E_{tot}^{Keating}$ based on the Keating potential are
compared to curvature energies ${\Delta}E_{C}^{Keating}$ based on
Keating-optimized geometries. (c) Strain energy ${\Delta}E$ in
different C$_{38}$ isomers. DFT total energies
${\Delta}E_{tot}^{DFT}$ are compared to curvature energies
${\Delta}E_{C}^{DFT}$ based on continuum elasticity theory for
structures optimized by DFT, and ${\Delta}E_{C}^{Keating}$ values
based on continuum elasticity theory for structures optimized
using the Keating potential. \label{fig3} }
\end{figure*}

\section{Validation of the continuum elasticity approach}\label{Sec3}

We will next test the accuracy of the continuum elasticity
approach by calculating the relative stability of non-planar
structures based on graphitic carbon. An infinite number of
morphologies including nanotubes, fullerenes and schwarzites may
be produced by deforming a segment of a graphene layer and
reconnecting its edges so that all carbon atoms are threefold
coordinated. In many cases, the non-planar structures contain
carbon pentagons and heptagons in the graphitic honeycomb
arrangement of atoms as required by Euler's
theorem.\cite{Encyclopedia-of-math}

To validate the continuum elasticity theory results, we calculated
the total energy of a graphene monolayer and selected graphitic
structures using {\em ab initio} density functional theory (DFT)
as implemented in the \textsc{SIESTA} code.\cite{SIESTA} We used
the Local Density Approximation (LDA)\cite{{CA},{PZ}} and
Perdew-Burke-Ernzerhof (PBE)\cite{PBE} exchange-correlation
functionals, norm-conserving Troullier-Martins
pseudopotentials\cite{Troullier91}, and a double-$\zeta$ basis
including polarization orbitals. The 1D Brillouin zone of
nanotubes was sampled by $16$~$k$-points and the 2D Brillouin zone
of graphene by $16{\times}16$~$k$-points.\cite{Monkhorst-Pack76}
The small Brillouin zones of schwarzites with several hundred C
atoms per unit cell
were sampled by only $1$~$k$-point. We used a mesh cutoff energy
of $180$~Ry to determine the self-consistent charge density, which
provided us with a precision in total energy of
${\alt}2$~meV/atom. All geometries have been optimized using the
conjugate gradient method,\cite{CGmethod} until none of the
residual Hellmann-Feynman forces exceeded $10^{-2}$~eV/{\AA}.

\subsection{DFT results for fullerenes}

Our DFT-LDA results for the relative energy
${\Delta}E_{tot}^{DFT}$ of optimized C$_n$
fullerenes\cite{{fowler1995atlas},{SM-fulsta14}} with respect to
graphene are shown in Fig.~\ref{fig3}(a). The various data points
for one size correspond to different structural isomers, which are
increasing fast in number with increasing $n$.
If all fullerenes were perfect spheres, Eq.~(\ref{eq4}) would
simplify to\cite{DT071} ${\Delta}E_{C}=4{\pi}D(1+\alpha)$. Using
the proper elastic constants for graphene\cite{DT071} $D=1.41$~eV
and $\alpha=0.165$, we would estimate ${\Delta}E_{C}=20.6$~eV for
all fullerenes independent of size. The numerical values for the
different optimized fullerene isomers in Fig.~\ref{fig3}(a) are
all larger, indicating that variations in the local curvature and
bond lengths cause a significant energy penalty.

\subsection{Comparison between computational approaches for C$_{38}$
fullerene isomers}

As we show in the following,
considering only local curvature variations across the surface
(and ignoring precise atomic positions) allows continuum
elasticity theory to quantitatively predict the strain energy with
a precision competing with {\em ab initio} calculations. To
illustrate this point, we present in Fig.~\ref{fig3}(b) the total
strain energy ${\Delta}E$ in seventeen isomers of C$_{38}$
obtained using various approaches. The strain energy
${\Delta}E_{tot}^{DFT}$ based on DFT, which is expected to
represent closely the experimental results, is not only
significantly lower than the predicted values
${\Delta}E_{tot}^{Tersoff}$ based on the Tersoff
potential,\cite{Tersoff88} but also differs from this popular
bond-order potential in the prediction of relative stabilities.

Next we demonstrate that accurate energy estimates may be obtained
even for geometries optimized using simple potentials with only
bond stretching and bond bending terms such as the Keating
potential\cite{{Keating66},{DT059}}
%
%
\begin{equation}
{\Delta}E_K = \frac{1}{2}\alpha_K \mkern-14mu
   \sum_{\stackrel{\scriptstyle <i,j>}{i<j}} \!\!
                     \frac{(r_{ij}^2 - R^2)^2}{R^2}
            + \frac{1}{2}\beta_K \mkern-18mu
   \sum_{\stackrel{\scriptstyle <i,j,k>}{j<k}} \mkern-12mu
                     \frac{({\bf{r}_{ij}} \cdot {\bf{r}_{ik}}
                           +\frac{1}{2}R^2)^2}{R^2} \; .
\end{equation}
The first term sums over nearest neighbor pairs and the second
term over nearest neighbor triplets, where $j$ and $k$ share the
same neighbor $i$. DFT calculations for graphene yield
$R=1.42$~{\AA} as bond length, $120^\circ$ as bond angle,
${\alpha}_K=11.28$~eV/{\AA}$^2$ and ${\beta}_K=4.14$~eV/{\AA}$^2$.

Geometries of C$_{38}$ fullerene isomers optimized by DFT and the
Keating potential are presented in the Supplemental
Material.\cite{SM-fulsta14} Strain energies for Keating optimized
fullerenes are shown in Fig.~\ref{fig3}(b). Whereas the Keating
optimized geometry is close to the DFT optimized geometry, the
Keating strain energy ${\Delta}E_{tot}^{Keating}$ clearly
underestimates the DFT values and does not correctly represent the
relative stabilities of the different isomers. As an alternative,
we used the Keating optimized geometry to obtain the curvature
strain energy ${\Delta}E_{C}^{Keating}$ using the continuum
approach. We found that this approach represents the relative
stabilities of isomers adequately and compares well to
${\Delta}E_{tot}^{DFT}$. The curvature strain energy values are
somewhat lower than the DFT values, since energy penalties
associated with bond stretching and bending do not appear in the
continuum approach. The small value of such corrections reflects
the fact that in equilibrated structures, bond lengths and angles
are near their optimum. The largest errors are expected in
frustrated structures, where not all bond lengths and angles can
be optimized simultaneously.


One of the key findings of this study is that continuum elasticity
theory provides not only a fast, but also a relatively robust way
to determine relative stabilities that are, to some degree,
insensitive to the precise geometry. We illustrate this point in
Fig.~\ref{fig3}(c), where we compare different ways to determine
the total strain energy ${\Delta}E$ in all C$_{38}$ isomers
discussed in Fig.~\ref{fig3}(b). ${\Delta}E_{tot}^{DFT}$, shown by
the solid line, is the difference between the total energy in DFT
of DFT-optimized C$_{38}$ isomers and 38 carbon atoms in the
graphene structure. ${\Delta}E_{C}^{DFT}$, given by the dashed
line, is the curvature energy based on the DFT-optimized geometry.
${\Delta}E_{C}^{Keating}$, given by the dash-dotted line, is the
curvature energy based on the Keating-optimized geometry. We note
that all expressions provide an accurate representation of
relative stabilities. As mentioned above, the fact that
${\Delta}E_{C}$ is about 10\% lower than ${\Delta}E_{tot}$ is
caused by our neglecting the stretching and bending of discrete
atomic bonds in the continuum approach.

\subsection{Comparison of computational approaches for $sp^2$
bonded carbon nanostructures}

\begin{figure}[!t]
\includegraphics[width=0.7\columnwidth]{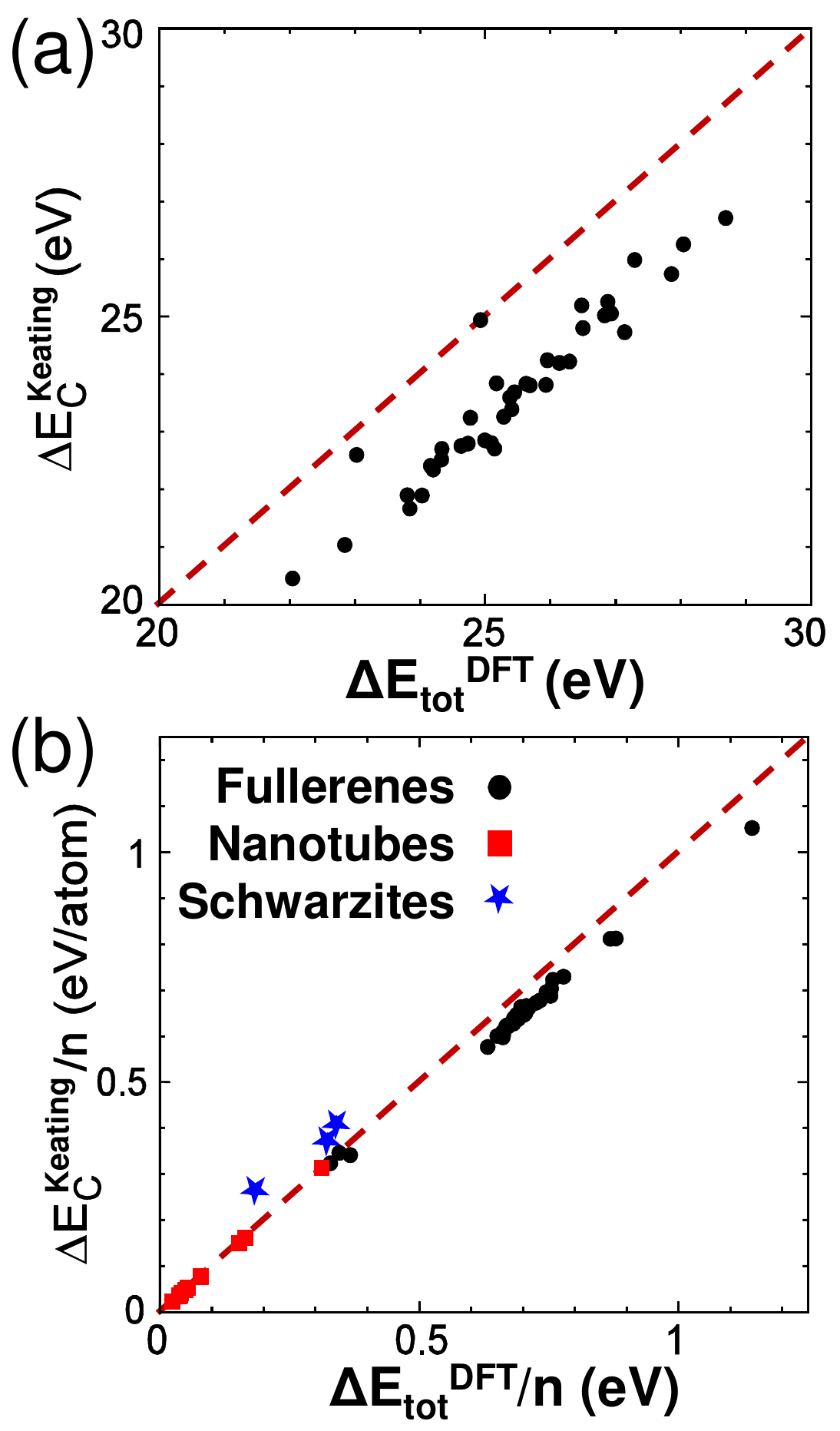}
\caption{%
(Color online) Strain energy ${\Delta}E$ in carbon nanostructures
with respect to the graphene reference system. (a) Comparison
between DFT-based total energies ${\Delta}E_{tot}^{DFT}$ and the
curvature energy ${\Delta}E_{C}^{Keating}$ based on
Keating-optimized geometries for all fullerene isomers considered
in Fig.~\protect\ref{fig3}(a). (b) Comparison between DFT-based
strain energies ${\Delta}E_{tot}^{DFT}/n$ and curvature energies
per atom ${\Delta}E_{C}^{Keating}/n$ for Keating-optimized
geometries of fullerenes, nanotubes and schwarzites. Dashed lines
represent agreement between DFT and continuum elasticity results. %
\label{fig4} }
\end{figure}

Encouraged by the level of agreement for C$_{38}$, we present in
Fig.~\ref{fig4}(a) the correlation between the curvature energy
${\Delta}E_{C}^{Keating}$
and ${\Delta}E_{tot}^{DFT}$ based on DFT for all fullerenes
discussed in Fig.~\ref{fig3}(a). The narrow spread of the data
points around the ${\Delta}E_{C}^{Keating}={\Delta}E_{tot}^{DFT}$
line confirms that the continuum elasticity approach is
competitive in accuracy with computationally much more involved
{\em ab initio} calculations.

To demonstrate the generality of our approach, we extend it from
near-spherical fullerenes to nanotubes with cylindrical symmetry
and schwarzites with local negative Gaussian curvature. Since
nanotubes and schwarzites are infinitely large, we compare
stabilities on a per-atom basis in these structures. Besides
results for the fullerenes discussed in Figs.~\ref{fig3} and
\ref{fig4}(a), Fig.~\ref{fig4}(b) displays results for
nanotubes with radii ranging between $2.5-9.0$~{\AA} and for
schwarzite structures with $152$, $192$ and $200$ carbon atoms per
unit cell. These results again indicate an excellent agreement
between curvature energies in Keating-optimized structures and
DFT-based strain energies.
This agreement is particularly impressive, since the spread of
atomic binding energies extends over more than 1~eV.

\subsection{Phosphorene nanotubes}

\begin{figure}[!tb]
\includegraphics[width=1.0\columnwidth]{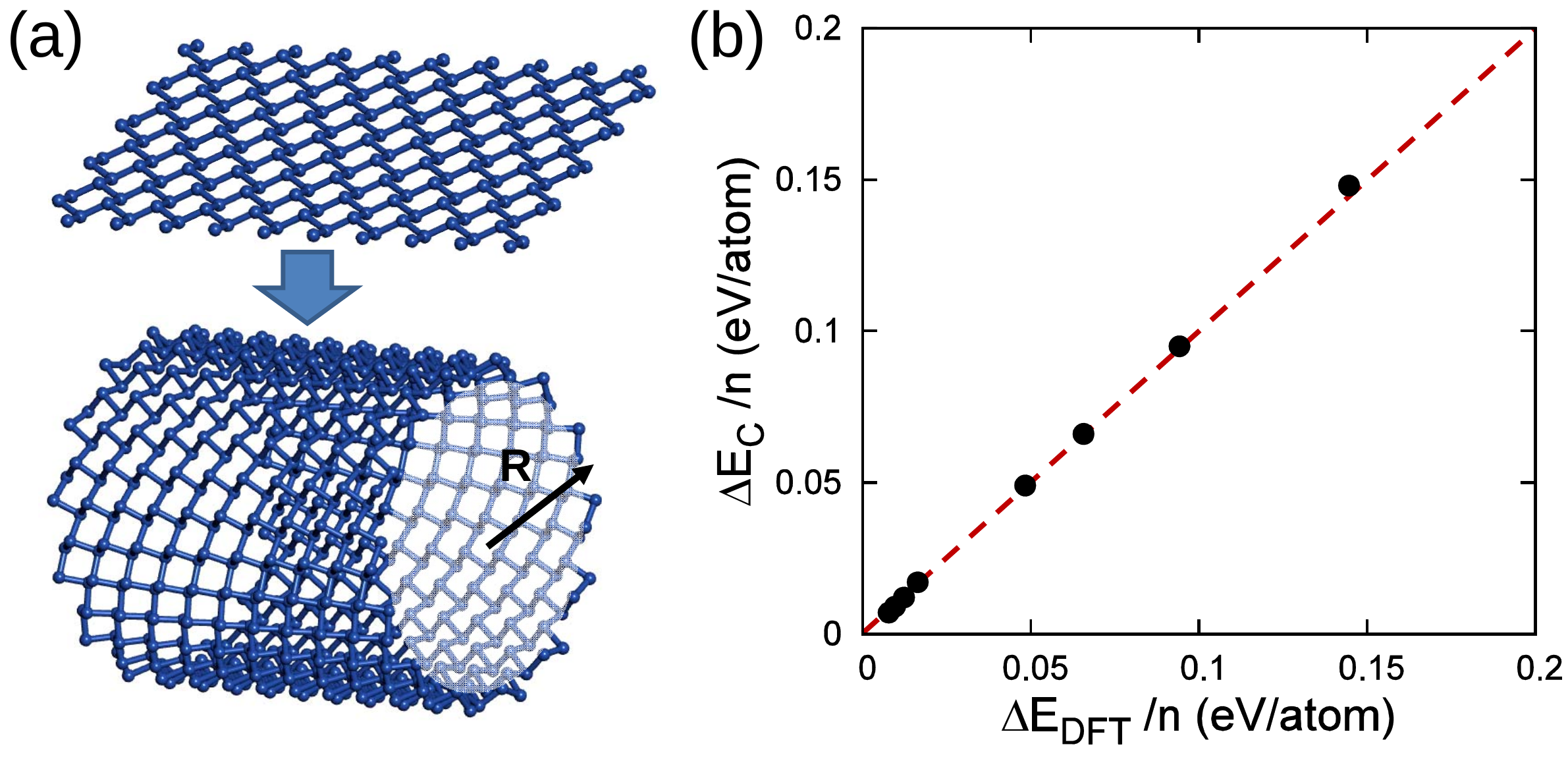}
\caption{%
(Color online) (a) Perspective view of the planar structure of a
blue phosphorene monolayer (top), which has been rolled up to a
nanotube with radius $R$ (bottom). (b) Comparison between the
strain energy per atom ${\Delta}E_{C}/n$ based on continuum
elasticity theory and ${\Delta}E_{tot}^{DFT}/n$ based on DFT in
blue phosphorene nanotubes. The dashed line represents agreement
between DFT and continuum
elasticity results.%
\label{fig5} }
\end{figure}

As suggested at the outset, our approach to estimate relative
stabilities is particularly valuable for unexplored systems such
as monolayers of blue phosphorus,\cite{DT230} where model
potentials have not yet been proposed. Our DFT-PBE results for a
blue phosphorene monolayer indicate $A=4.78$~{\AA}$^2$ as the
projected area per atom, $D=0.84$~eV and $\alpha=0.10$. The
monolayer structure, shown in the top panel of Fig.~\ref{fig5}(a),
has an effective thickness of 1.27~{\AA}. This structure can be
rolled up to phosphorene nanotubes with different radii $R$
using the approach used in the construction of carbon
nanotubes.\cite{DT227} As seen in Fig.~\ref{fig5}(b), the strain
energy for this geometry, obtained using continuum elasticity
theory, agrees very well down to very small radii with results
obtained using much more involved DFT calculations.\cite{PNT14}

\section{Discussion}


Given a set of points in space, such as atomic positions, it is
possible to construct a smooth surface that contains all these
points in order to characterize its shape everywhere, and to
eventually determine the deformation energy using the continuum
elasticity approach.

\begin{figure}[!t]
\includegraphics[width=0.8\columnwidth]{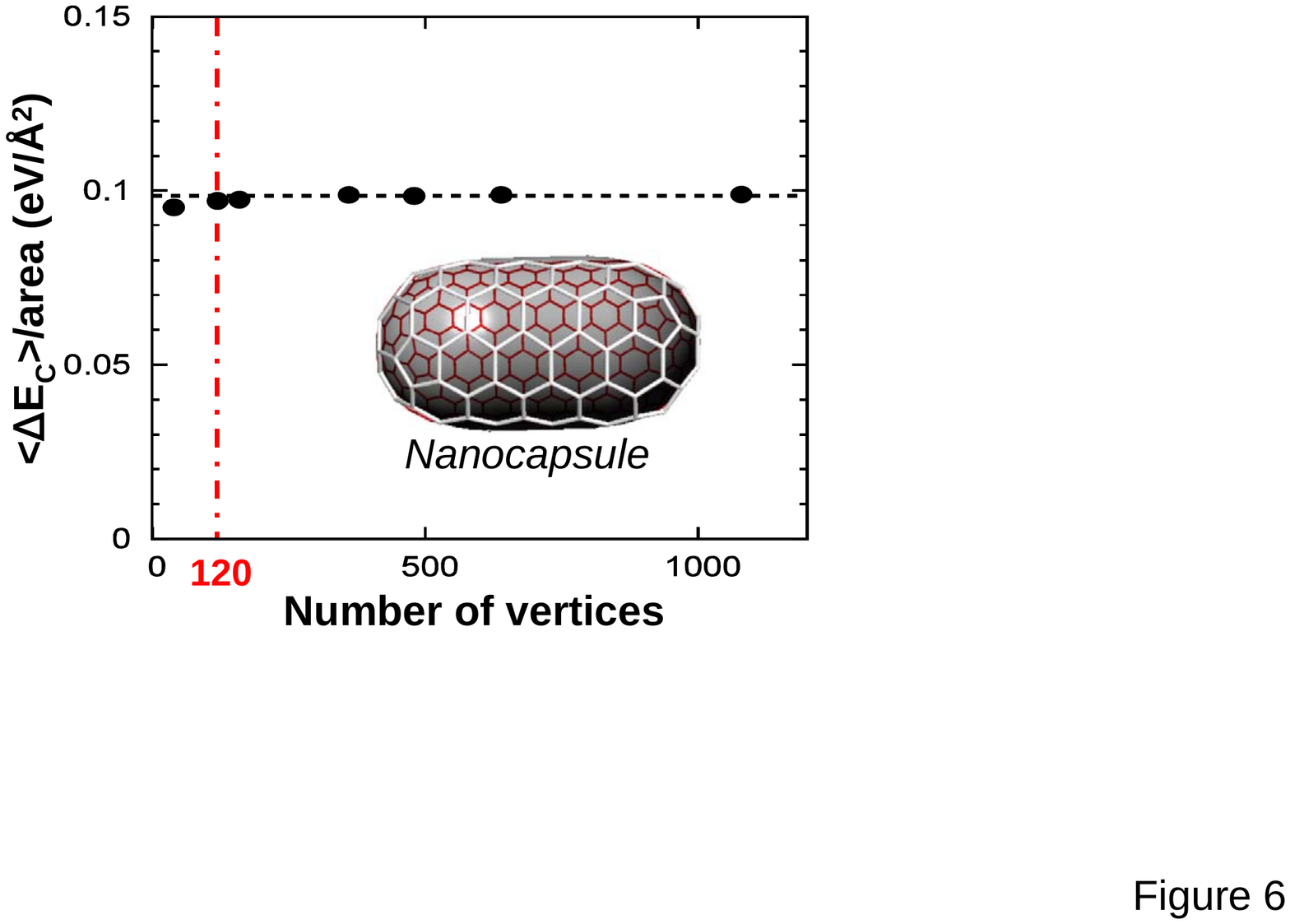}
\caption{%
(Color online) Average curvature energy $<{\Delta}E_C>$ per area
of a nanocapsule tessellated by a honeycomb lattice with different
numbers of vertices. The vertical dash-dotted red line indicates
that the capsule represents the C$_{120}$ structure. The inset
shows, how the capsule surface can be tessellated by a honeycomb
lattice with 120 vertices or atoms, shown by the white lines, and
also with 480 vertices, shown by the red lines. The horizontal
dashed black line represents an extrapolation to an infinitely
dense tessellation. \label{fig6} }
\end{figure}

We illustrate this point by tessellating the smooth surface of a
graphitic nanocapsule, consisting of a cylinder capped by
hemispheres at both ends and representing C$_{120}$, in different
ways. Our results in Fig.~\ref{fig6} show that the average
curvature energy $<{\Delta}E_C>$ is rather insensitive to the
tessellation density. The horizontal dashed line at
$<{\Delta}E_C>=0.099$~eV/{\AA}$^2$,
representing an extrapolation to a dense tessellation, is
${\approx}5$\% higher than the exact continuum elasticity value of
$0.093$~eV/{\AA}$^2$,
obtained for an ideal capsule with cylinder and hemisphere radius
$R=3.55$~{\AA}.
The small difference arises from our approximate
way to estimate the mean curvature $k$ on the cylinder surface and
at the interface between the cylinder and the hemisphere. The
extrapolated value is also close to the
$<{\Delta}E_{tot}^{DFT}>=0.100$~eV/{\AA}$^2$ based on the
DFT-optimized C$_{120}$ capsule.

The reverse process to determine atomic positions from the shape
alone is not unique. An informative example is the structure of a
carbon nanotube. Whereas the precise atomic structure within each
nanotube is defined by the chiral index, many nanotubes with
different chiral indices share essentially the same diameter and
the same local curvature. Thus, given only the diameter of a
(wide) hollow cylinder representing a nanotube, it is impossible
to uniquely identify the chiral index and thus the atomic
position. As a matter of fact, identifying the precise atomic
positions is not necessary, since according to continuum
elasticity theory, supported by experimental evidence, the
stability of nanotubes depends only on the tube
diameter.\cite{DT227}

From its construction, the continuum elasticity description of
local and global stability is best suited for very large
structures with small local curvatures.
Therefore, the high level of agreement between its predictions and
{\em ab initio} results in structures with large local curvatures
is rather impressive. Among the different allotropes, we find the
continuum elasticity description to be most accurate for carbon
nanotubes, where all bond lengths are at their equilibrium value.
In fullerenes and schwarzites, the presence of non-hexagonal
rings, including pentagons and heptagons, prevents a global
optimization of bond lengths and bond angles, reducing the
agreement with DFT results.

Our stability results are consistent with the pentagon adjacency
rule that provides an energy penalty of $0.7-0.9$~eV for each pair
of adjacent pentagons,\cite{{Zhang92},{Warshel72},{Negri88}} which
causes an increase of the local curvature. While this rule is
surely useful, it can not compare the stability of isomers with
isolated pentagons or structures of different size.

What we consider the most significant benefit of our approach to
determine local strain\cite{SM-fulsta14} is to identify the least
stable sites in a structure. Local curvature and in-plane strain
play the key role in both local stability and local electronic
structure,\cite{Pereira09} which also controls the chemical
reactivity.\cite{{Lu2005cr},{cnt-deformation}} Thus, our approach
can identify the most reactive and the least stable sites, which
control the stability of the entire system.\\

\section{Summary and conclusions}

In conclusion, we have introduced a fast method to determine the
local curvature in 2D systems with arbitrary shape. The curvature
information, combined with elastic constants obtained for a planar
system, provides an accurate estimate of the local stability in
the framework of continuum elasticity theory. Relative stabilities
of graphitic structures including fullerenes, nanotubes and
schwarzites, as well as phosphorene nanotubes calculated using
this approach, agree closely with {\em ab initio} density
functional calculations. The continuum elasticity approach can be
applied to all 2D structures and is particularly attractive in
complex systems with known structure, where the quality of
parameterized force fields has not been established.\\

\begin{acknowledgments}
We acknowledge valuable discussions with Zacharias Fthenakis. This
work was funded by the National Science Foundation Cooperative
Agreement \#EEC-0832785, titled ``NSEC: Center for High-rate
Nanomanufacturing''. Computational resources for this project were
provided by the Michigan State University High-Performance
Computer Center.
\end{acknowledgments}


%

\end{document}